\documentclass[a4paper,10pt,twoside]{cpc-hepnp}
\usepackage{subfigure}
\usepackage{mathrsfs}
\usepackage{multicol}
\usepackage{graphicx}
\usepackage{booktabs}
\usepackage{amssymb,bm,mathrsfs,bbm,amscd}
\usepackage[tbtags]{amsmath}
\usepackage{lastpage}
\usepackage{amsmath, amssymb}
\usepackage{slashed}
\usepackage{multirow}
\usepackage{CJK}
\usepackage{lineno}
\usepackage{color}

\begin{document}
\begin{CJK*}{GBK}{song}

\fancyhead[c]{\small Chinese Physics C~~~Vol. xx, No. x (2022) xxxxxx} \fancyfoot[C]{\small xxxxx-\thepage}

\footnotetext[0]{Received \today}

\title{\boldmath Classify the Higgs decays with the PFN and ParticleNet at electron-positron colliders}

\author{Gang Li$^{1,1)}$\email{li.gang@ihep.ac.cn},
\quad Libo Liao$^{2,2)}$\email{liaolb-jay@outlook.com}, 
\quad Xinchou Lou$^{1}$,
\quad Peixun Shen$^1$
\quad Weimin Song$^3$, \\
\quad Shudong Wang$^{1,4,3)}$\email{wangsd@ihep.ac.cn}, 
\quad and Zhaoling Zhang$^3$}
\maketitle

\address{
$^1$ Institute of High Energy Physics, Chinese Academy of Sciences,  19B Yuquan Road, Shijingshan District, Beijing, China\\
$^2$ Wuzhou University, 82 Fumin Third Road, Wanxiu District, Wuzhou, China,\\
$^3$ College of Physics, Jilin University, 2699 Qianjin Street, Changchun, China,\\
$^4$ School of Physical Sciences, University of Chinese Academy of Sciences, Beijing, China
}

\abstract{
  Various Higgs factories are proposed to study the Higgs boson precisely and systematically in a model-independent way.  In this study, the Particle Flow Network and ParticleNet techniques are used to classify the Higgs decays into multi-categories and the ultimate goal is to realize an ``end-to-end'' analysis. A  Monte Carlo simulation study is performed to demonstrate the feasibility, and the performance looks rather promising. This result could be the basis of a ``one-shop" analysis to measure all the branching fractions of the Higgs decays simultaneously.   
}


\begin{multicols}{2}
\section{Introduction}

The historic observation of the Higgs boson in 2012 at the Large Hadron Collider (LHC)~\cite{higgs_atlas,higgs_cms} declared the discovery of the last missing piece of the most fundamental building blocks in the Standard Model (SM). Although the SM has been remarkably successful in describing experimental phenomena, a precision Higgs physics program would be critically important given that the SM does not predict the parameters in the Higgs potential, nor does it involve particle candidates for dark matter. The precision determination of the Higgs couplings to the SM particles, gauge bosons and leptons/quarks, are the agents probing the Higgs mechanism for generating the masses~\cite{couplings}. In particular, potential observable deviations of the Higgs couplings from the SM expectations would indicate new physics~\cite{newphysics}. Therefore, the Higgs discovery marks the beginning of a new era of both theoretical and experimental explorations. Various $e^+e^-$ colliders as Higgs factories were proposed by the high energy physics community, such as ILC~\cite{Baer:2013cma}, CLIC~\cite{Aicheler:2012bya}, FCC-ee~\cite{Abada:2019zxq}, and CEPC~\cite{CDR-A, CDR-D}.

The most important advantages of a Higgs factory are that the center of mass (CM) energy is precisely defined and that they could perform absolute measurements to the Higgs boson. Neglecting the $Z$ fusion production, in an $e^+e^-\to ZH$ event, where the $Z$ decays to a pair of visible fermions ( $Z\rightarrow e^+e^-,~\mu^+\mu^-,~\tau^+\tau^-,~\mbox{or}~q\bar{q}$ ), the Higgs boson can be identified with the kinematics of these fermion pairs independent of its decays. The production cross-section and most of the decay branching fractions of the Higgs could be measured model-independently by the counting method. For example, the CEPC can measure the cross-section of $e^+e^-\rightarrow ZH$, $\sigma(ZH)$, at 240\,GeV, to a precision of 0.5\% and the branching fractions to a few percent, respectively, by combining the four decay modes of the $Z$ boson~\cite{CDR-D,an2019}. 

The physics goal of a Higgs factory must be accomplished by optimizing the detector design and making use of the latest developments in data science. Recently, various Machine Learning (ML) techniques have already shown very promising performances in the data analysis of high energy physics because of their strong inductive biases\cite{Schwartz:2021ftp}, in particular for jet studies. For instance, jets are treated as images~\cite{jetImage01,jetImage02,jetImage03,jetImage04,jetImage05,jetImage06} or as sequences~\cite{sequence01,sequence02,sequence03,sequence04}, trees~\cite{tree01, tree02}, graphs~\cite{graph01}, or sets~\cite{EFN,particlenet} of particles, and ML techniques, most notably deep neural networks (DNNs), are used to build new jet tagging algorithms automatically from (labeled) simulated samples and even data~\cite{DNN01, DNN02, DNN03, DNN04}.  While the above ML techniques are used at jet-level for case studies, they naturally can apply for the event level in $e^+e^-$ collision, which has much simpler topology and pile-up free.

In this article, two ML approaches are used studying the classification problem of Higgs events. The classification results can serve as the basis of an ``end-to-end" (E2E) analysis, which means that it starts from particle level information  and is efficient and balanced because it can study almost all the Higgs decays modes simultaneously through state-of-the-art ML techniques. The approach also is a "one-shop" analysis to support extracting all Higgs couplings and taking into account the correlations and commonalities of the same detector for the experiment. 

The rest of this paper is organized as follows. The ML methods used in this study are introduced in Sec.\,2, followed by the implementation of the ML with a Monte Carlo (MC) simulation in Sec.\,3. Finally, a summary is presented.

\section{Machine Learning methods}

Recently, various ML techniques were proposed for jet tagging studies. Among them, the PFN~\cite{EFN} and ParticleNet~\cite{particlenet} got relatively superior performance.  

In~\cite{EFN} the authors applied the Deep Sets concept~\cite{deepsets} to the jet tagging problem. They propose two elegant model architectures, named EnergyFlow Network (EFN) and ParticleFlow Network (PFN), with provable physics properties, such as infrared and colinear safety. In these two architectures, the features of each particle are encoded into a latent space of $\Phi$~\cite{deepsets} and the category, $F$, is extracted from the summed representation in that latent space.  Both $\Phi$ and $F$ are approximated by neural networks. The key mathematical fact is that a generic function of a set of particles can be decomposed into an arbitrarily good approximation according to the {\bf Deep Set Theorem}~\cite{deepsets}.  The performance of these models in classification problems is comparable with other more complicated models. The authors also tried to interpret and visualize what the model has learned~\cite{EFN}.

Motivated by the success of CNNs, the ParticleNet~\cite{particlenet} approach based on the Dynamic Graph Convolutional Neural Network (DGCNN) is proposed for learning on particle cloud data. The edge convolution (``EdgeConv") operation, a convolution-like operation for point clouds, is used instead of the regular convolution operation.  One important feature of the EdgeConv operation is that it can be easily stacked, just as regular convolutions.  Therefore, another EdgeConv operation can be applied subsequently, which make it possible to learn features of point clouds hierarchically. Another important feature is that the proximity of points can be dynamically learned with EdgeConv operations. The study shows the graph describing the point clouds are dynamically updated to reflect the changes in the edges, i.e., the neighbors of each point. Reference~\cite{particlenet} shows that this leads to better performance than keeping the graph static.

As suggested by the authors~\cite{EFN} and according to the performances of EFN and PFN, we choose the PFN and ParticleNet to classify the Higgs decays. This ML attempt contains some distinct features in contrast to conventional data analysis. First, the ML approach is used to classify  many physics processes at the same time. If some tiny decays are neglected, there are about 9 branching fractions of the Higgs decay to be measured. The number of classes is greater than 9 when the SM backgrounds are included.  In addition, the classification results could be the basis of an E2E analysis, which means that all the particle level information, such as four momenta, PID, and impact parameters of charged particles, is used as input directly, and the network calculates the scores of each event.  In this case, the analysis no longer needs some dedicated and complicated reconstruction tools, such as lepton/photon isolation, jet-clustering, and $\tau$ finder, and so on.

\section{Classify the Higgs decays}
In this section, a 9-classification of the all accessible Higgs decay final states is realized with the PFN method, and the confusion matrices of them are determined. And as an attempt, a more ambitious 39-classification is tried with the ParticleNet, and promising and consistent results are achieved.

\subsection{ML model setup}
For the two ML models used to classify the Higgs decays, kinematic information of energy, polar and azimuthal angles are always given for each particle. We should note that energies and polar angles are used instead of the transverse momenta and rapidities in the original study~\cite{EFN, particlenet}, respectively, since the models are utilized for $e^+e^-$ collider experiments in this study. All features are pre-processed: the energies are divided by the scalar $\sum E$ and the polar and azimuthal angles are centered based on those of the event. The inputs also include the PID and impact parameters of charged particles.

The PFN architecture~\cite{EFN} is designed to parameterize the functions $\Phi$ and $F$ in a sufficiently general way, several dense neural network layers are used as universal approximators. For $\Phi$, three dense layers are employed with 100, 100, and $l$ nodes, respectively, where $l$ is the latent dimension that takes 256 after comparing the performances of 128 and 256.  For $F$, we use the same configuration as the original paper, which has three dense layers, each with 100 nodes. Each dense layer uses the {\rm ReLU} activation function and He-uniform parameter initialization~\cite{he-uniform}. A nine-unit layer (depends on the number of classes) with a {\rm SoftMax} activation function is the output layer. 

The ParticleNet\cite{particlenet} architecture consists of three EdgeConv blocks, one aggregation layer, and two fully-connected layers. The first EdgeConv block uses the spatial coordinates of the particles in the $\theta-\phi$ space to compute the distances, while the subsequent blocks use the learned feature vectors as coordinates. The number of nearest neighbors $k$ is 16 for all three blocks, and the number of channels $C$ for each EdgeConv block is (64, 64, 64), (128, 128, 128), and (256, 256, 256), respectively. After the EdgeConv blocks, a channel-wise global average pooling operation is applied to aggregate the learned features over all particles in the cloud. This is followed by a fully connected layer with 256 units and the ReLU activation. A dropout layer with a drop probability of 0.1 is included to prevent overfitting. A fully connected layer with 39 units, followed by a softmax function, is used to generate the output for the 39 classification task.

\subsection{Data samples}
In this study, there are 4 production modes of the Higgs boson at 240 GeV to be analyzed, i.e.,  $e^+e^- \to e^+e^-H$, $\mu^+\mu^-H$, $\tau^+\tau^-H$, and $q\bar{q}H$. And in each production mode, the same 9 decay modes are measured,  which are
$H \to c\bar{c}$, $b\bar{b}$, $\mu^+\mu^-$, $\tau^+\tau^-$, $gg$, $\gamma\gamma$, $ZZ$, $W^+W^-$, and $\gamma Z$, respectively. So there are 36 processes in total.  For each process, 400,000 events are generated with the {\sc WHIZARD 1.9.5}~\cite{whizard} and fed to Pythia6~\cite{pythia6} for hadronization. It should be noticed that the sequential decays of $W$ and $Z$ are not dealt specifically to avoid the complication, though it can enhance the classification performance if more decay knowledge is used.  

All the generated samples are simulated by a simplified way to model detector responses. In detail, all particles are simulated according to the performance of the baseline detector in the {\it CEPC CDR}~\cite{CDR-D}. The momentum resolution of charged tracks 
is $\frac{\sigma(p_t)}{p_t} = 2\times 10^{-5}\oplus \frac{0.001} {p\sin^{3/2}\theta} [\mathrm{GeV}^{-1}]$, the energy resolution of photons
is $\frac{\sigma(E)}{E} = 0.01 \oplus \frac{0.20}{\sqrt{E/\mathrm{(GeV)}}}$, and this of neural hadrons is $\frac{\sigma(E)}{E} = 0.03 \oplus \frac{0.50}{\sqrt{E/\mathrm{(GeV)}}}$, which means all the particles are reconstructed with a perfect particle flow detector. In case of impact parameters and particle identification, they are taken directly from the truth of generation.  While the simulation is a bit too ideal, it is sufficient for a feasibility study.  

\subsection{9-category classification: training and evaluation }

\begin{figure}[htbp]
\centering 
\includegraphics[width=0.5\textwidth, origin=c,angle=0]{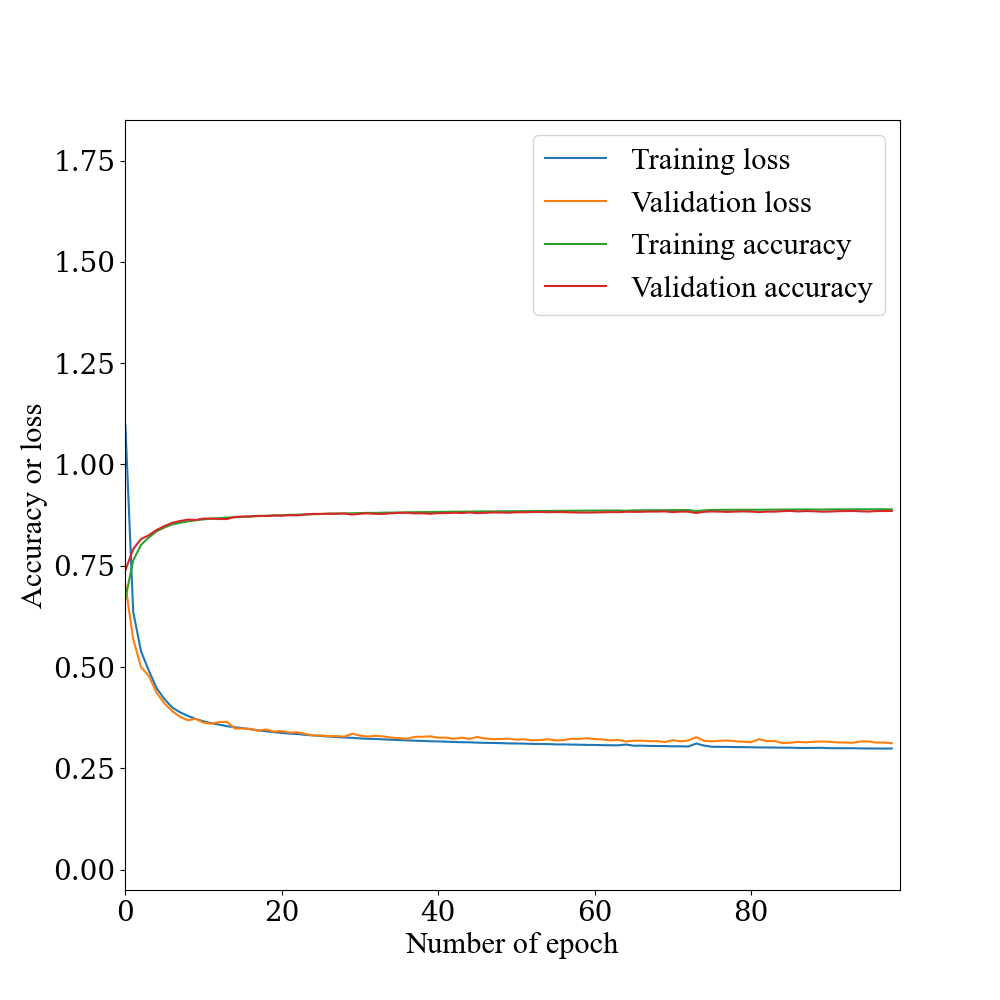}
\caption{\label{fig:loss} Accuracy and loss versus the number of epochs of $e^+e^- \to e^+e^- H$ process during the training. }
\end{figure}

During model training, the common properties of the neural network include categorical cross-entropy loss function, the Adam optimization algorithm~\cite{adam}, batch size of 1000, and a learning rate of 0.001. 400,000 events are used for each production mode, and the total number of events of 9 decays is 3,600,000. The full data set is split into training, validation, and test samples according to fractions of 8:1:1.  The monitoring of loss and accuracy on training and validation samples shows that the models converge well and there is no obvious over-training after the models are trained for 100 epochs, see Figure~\ref{fig:loss} as an example.
\end{multicols}

\begin{figure}[htbp]
\centering 
\includegraphics[width=.8\textwidth,origin=c,angle=0]{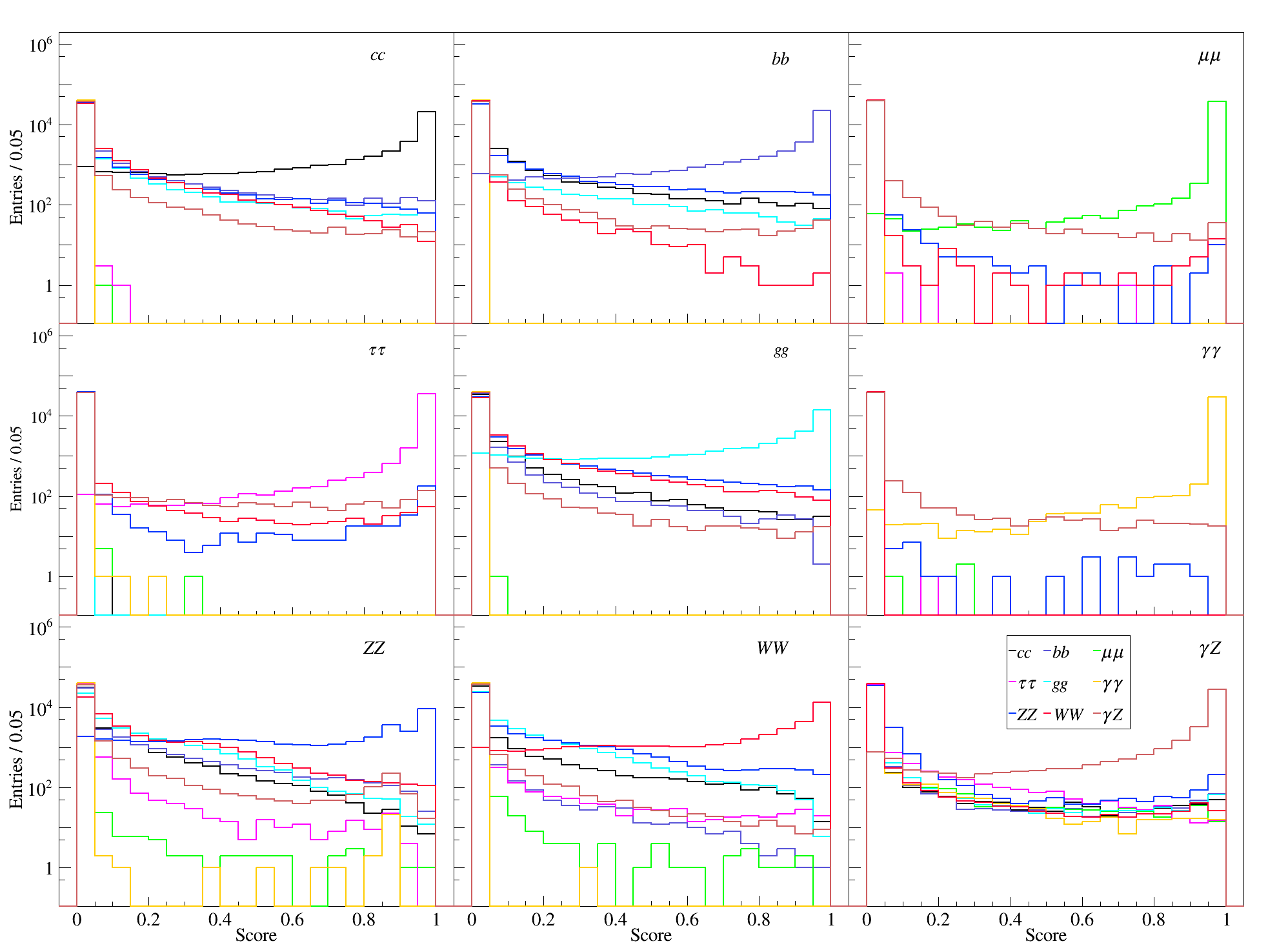}
\caption{\label{fig:softmax} The distributions of 9 outputs for each true category, taking $e^+e^-H$ as an example. Each score is calculated by assuming that the event belongs to that category. }
\end{figure}

\begin{figure}[htbp]
\centering 
\includegraphics[width=.8\textwidth,origin=c,angle=0]{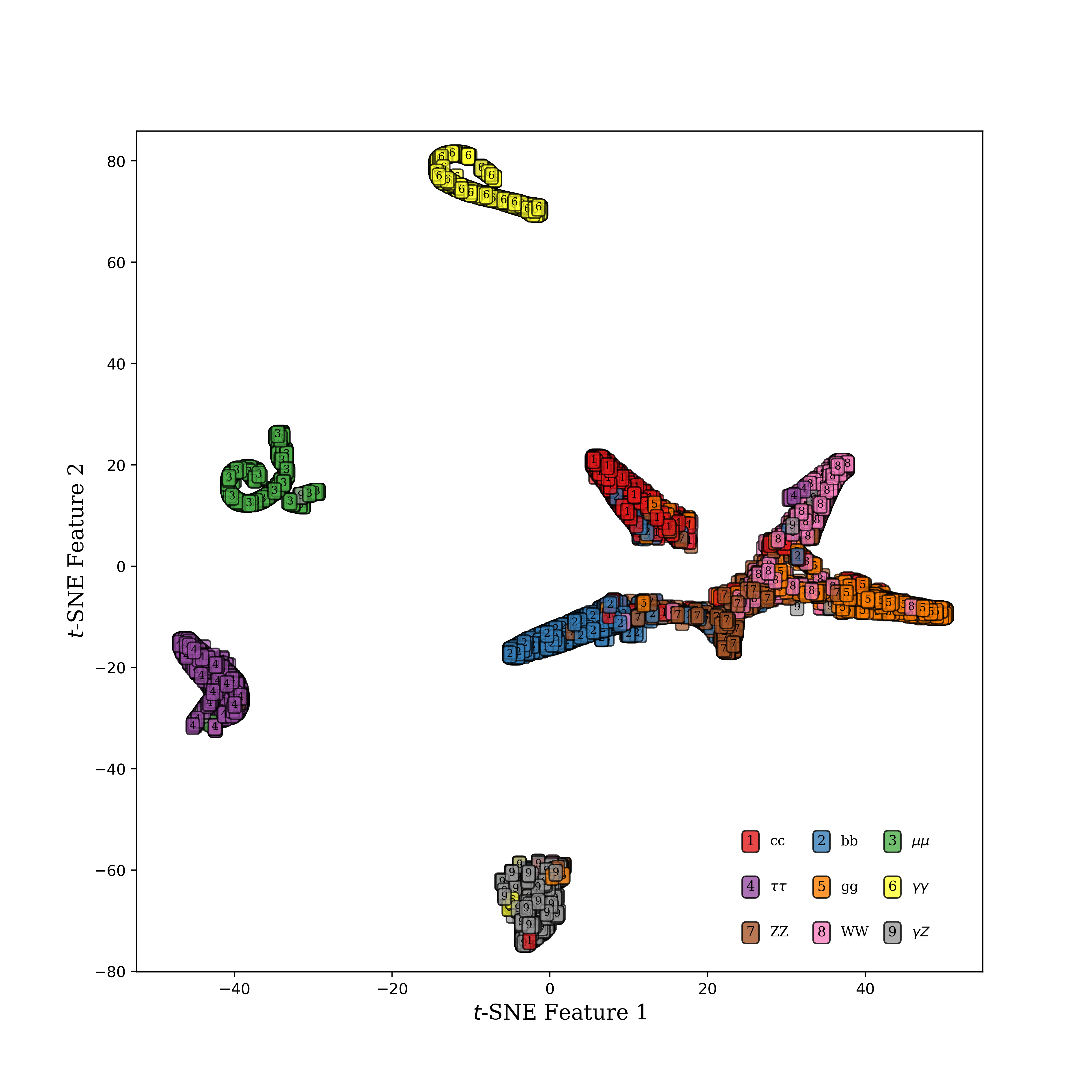}
\caption{\label{fig:tsne}
 The classification performance on the test set is visualized with $t$-SNE , where the two largest components are used, taking 10,000 events of $e^+e^-\to e^+e^-H$ for illustration.
}
\end{figure}

\begin{multicols}{2}
The outputs of the classifier, which are from a nine-unit layer with a {\rm SoftMax} activation function, are visualized in various ways.  The {\rm SoftMax} is essential because it helps to produce scores comprising 9 probabilities proportional to the exponential of the input information for each event, which is input for a cut-based data analysis. Figure~\ref{fig:softmax} presents the 9 scores for each category. Taking the bottom left panel as an example, these events are of $H\to ZZ$, and the curves in different colors represent the probability distributions assuming them to be $H\to ZZ$. The blue curve peaks when the score approaches 1, which means the classifier can identify  $H\to ZZ$ signals. There are two small peaks in the blue and brown curves around 0.8, which shows that $H\to Z Z$ and $H\to \gamma Z$ can contaminate each other because of the similarity of their cascade decays. From Figure~\ref{fig:softmax}, it can be seen that high-dimensional data is difficult to visualize intuitively.  A better way is that data in lower dimensions are plotted to show the inherent structures. To aid visualization of the structure of 9 outputs, the $t$-SNE\cite{tsne} method is used. Figure~\ref{fig:tsne} shows the distribution of the two largest components after the dimensionality reduction, where 1-9 represents the 9 decay modes of the Higgs boson from $c\bar{c}$ to $\gamma Z$ in the same order as the above, respectively. The patterns in Figure~\ref{fig:tsne} are consistent with those in Figure~\ref{fig:softmax} but much clearer. It can be seen that $\mu^+\mu^-$(3), $\gamma\gamma$(6), $\tau^+\tau^-$(4) and $\gamma Z$(9) modes are almost isolated clusters and background free. The clusters of the others can also be seen and the overlaps are also significant. 
\end{multicols}

\begin{multicols}{2}
Some standard quantities can measure the performances of classifiers. 
For instance,  accuracy (ACC) measures the fraction of correctly classified observations,   ROC curve (Receiver Operating Characteristic curve) visualizes the True Positive Rate (TPR) versus the False Positive Rate (FPR), and AUC (Area Under the Curve) is the area under the ROC curve. If we have a better classification for each threshold value, the area grows, and a perfect classification leads to an AUC of 1.0. The average ACC and AUC for all 36 processes in the 4 tagging modes are summarized in Table~\ref{tab:training}. Several conclusions can be drawn from the table. First, the average accuracy of each tagging mode reaches about 87\%, which is good and adequate for further analysis.  The decays of $H\to \mu^+\mu^-$, $\tau^+\tau^-$, and $\gamma\gamma$ have the best accuracy and largest AUCs as expected. Last but not least, the accuracy of $H\to ZZ$ or $W^+W^-$ is not good as the others, which leaves room for further improvement. 

\begin{figure}[htbp]
\centering 
\subfigure[]
{
\begin{minipage}[b]{0.47\textwidth}
\includegraphics[width=\textwidth,origin=c,angle=0]{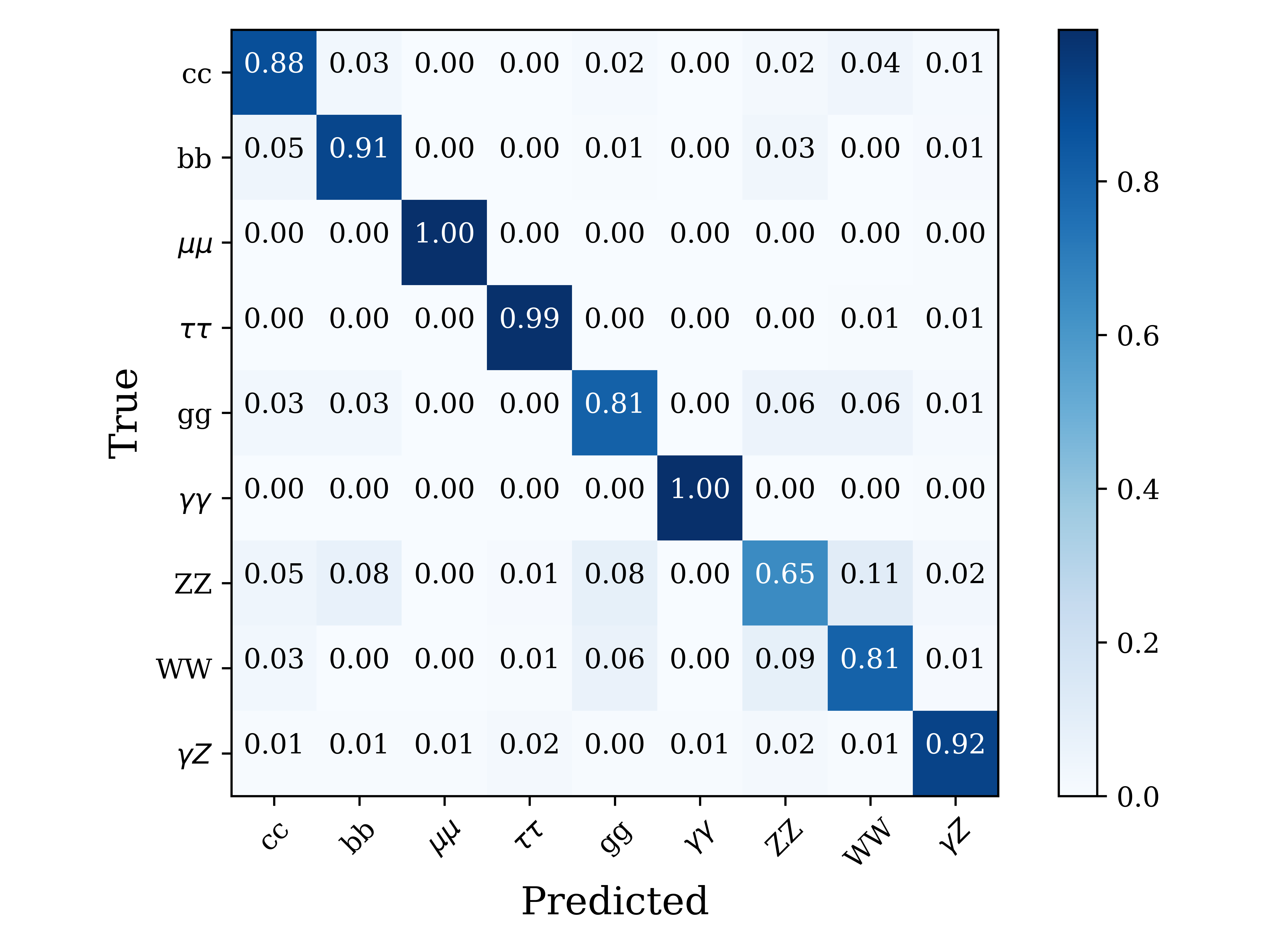}
\end{minipage}
}
\subfigure[]
{
\begin{minipage}[b]{0.47\textwidth}
\includegraphics[width=\linewidth,origin=c,angle=0]{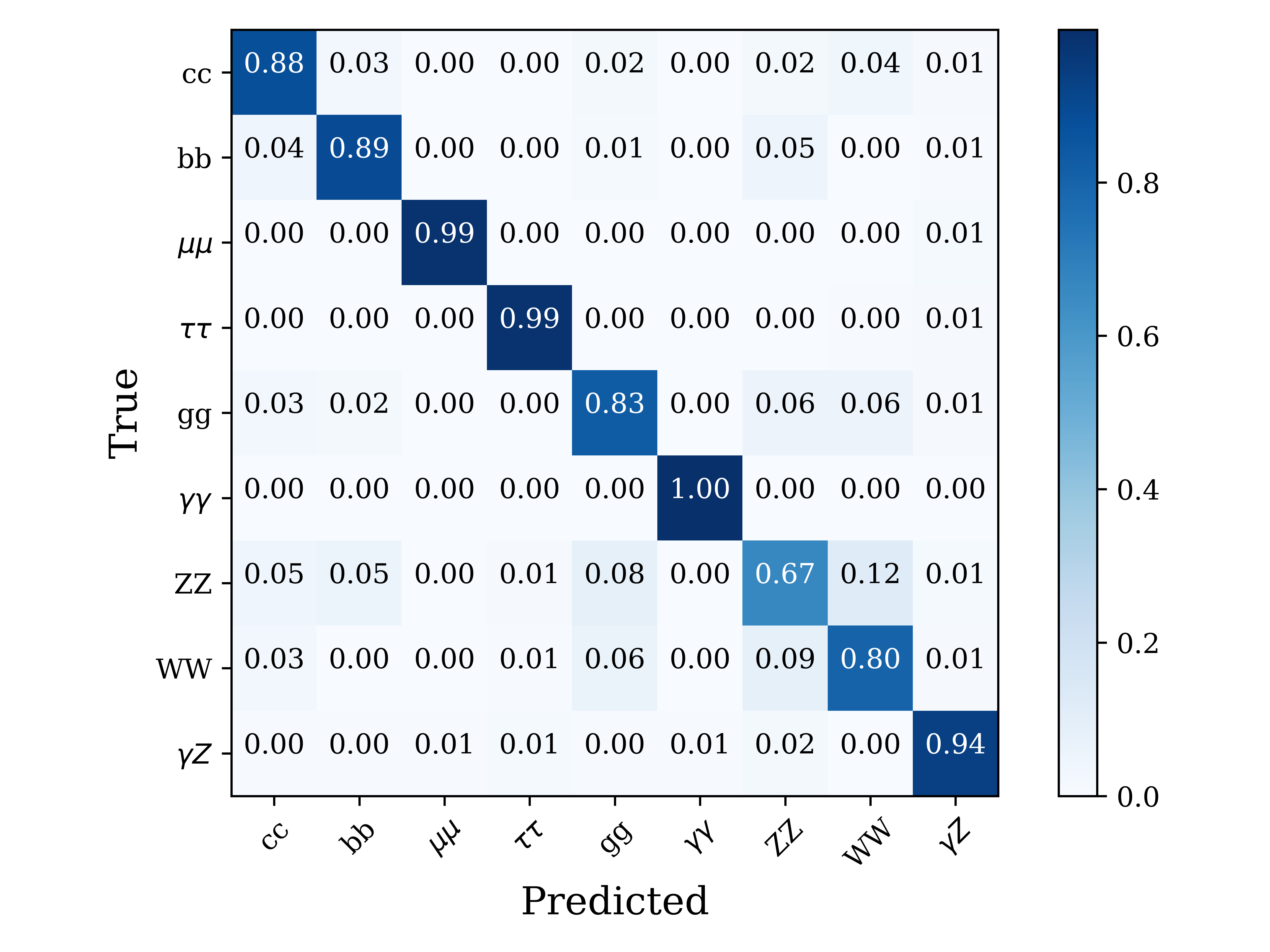}
\end{minipage}
}

\vfill

\subfigure[]
{
\begin{minipage}[b]{0.47\textwidth}
\includegraphics[width=\linewidth,origin=c,angle=0]{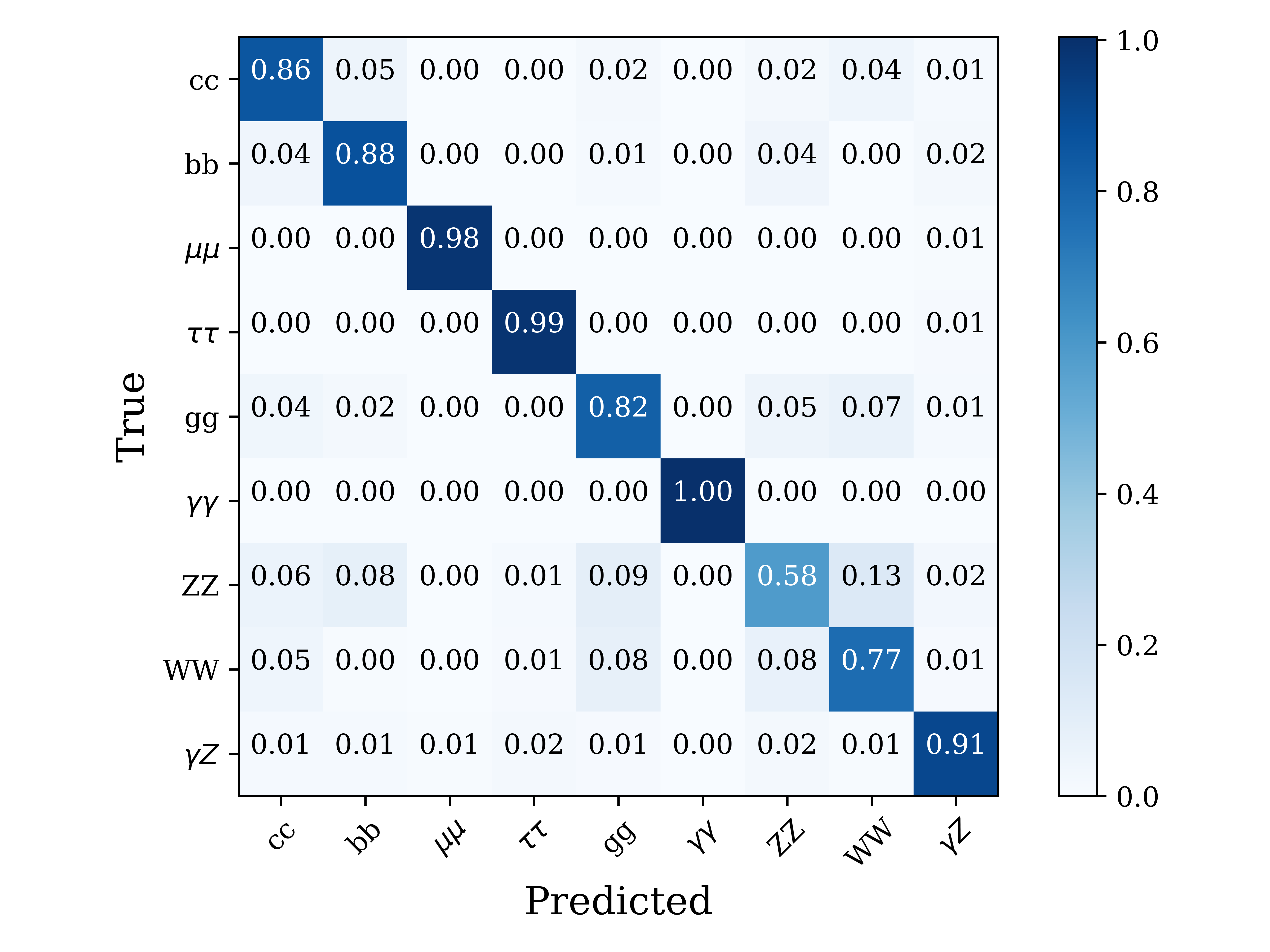}
\end{minipage}
}
\subfigure[]
{
\begin{minipage}[b]{0.47\textwidth}
\includegraphics[width=\linewidth,origin=c,angle=0]{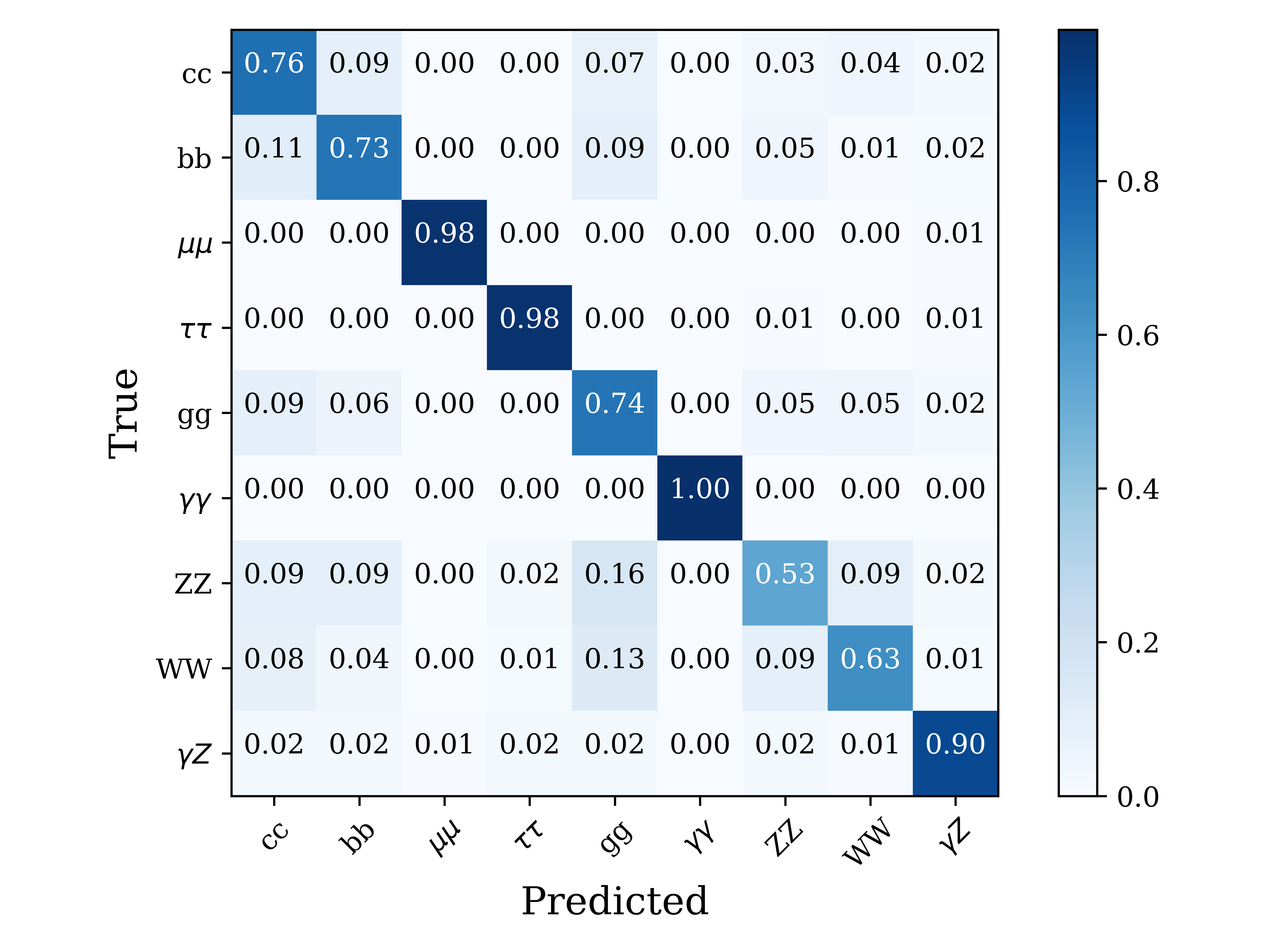}
\end{minipage}
}

\caption{\label{confusion} Confusion matrices of (a): $e^+e^- \to e^+e^- H $, (b): $e^+e^- \to \mu^+\mu^-H $, (c): $e^+e^- \to \tau^+\tau^-H $, and (d): $e^+e^- \to q\bar{q}H $, respectively.}
\end{figure}

Finally, the confusion matrices are used to evaluate the performance of the ML model and to be used as an important gradient for the further data analysis. Confusion matrices are calculated by comparing the prediction of the model and the true labels. Figure~\ref{confusion} shows the confusion matrices of the four classifiers, respectively. In terms of the confusion matrix, the accuracy appears as the diagonal elements of the corresponding confusion matrices, and the off-diagonal ones represent misclassification rates. So confusion matrices contain the complete information of both the correct and incorrect classifications, which could help to unfold the generated numbers of signals, $N_i$.  
\end{multicols}

\begin{table}[htbp] 
    \centering
    \begin{tabular}{l|cc|cc|cc|cc} 
    \toprule 
    \multirow{2}{*}{Decay Mode} \qquad  & \multicolumn{2}{c|}{$e^+e^-H$}  & \multicolumn{2}{c|}{$\mu^+\mu^- H$} & \multicolumn{2}{c|}{$\tau^+\tau^- H$}& \multicolumn{2}{c}{$ q\bar{q}H$} \\ 
                        \qquad  & ACC & AUC     \qquad &   ACC & AUC   \qquad  & ACC & AUC   \qquad & ACC & AUC    \\
    \midrule 
    $H\to c\bar{c}$     \qquad  & 0.880&0.991   \qquad &  0.882&0.991  \qquad  & 0.857&0.987 \qquad & 0.755&0.966  \\ 
    $H\to b\bar{b}$     \qquad  & 0.908&0.994   \qquad &  0.893&0.994  \qquad  & 0.877&0.991 \qquad & 0.733&0.972  \\ 
    $H\to \mu^+\mu^-$   \qquad  & 0.997&1.000   \qquad &  0.986&1.000  \qquad  & 0.981&1.000 \qquad & 0.983&1.000  \\ 
    $H\to \tau^+\tau^-$ \qquad  & 0.993&0.999   \qquad &  0.985&0.999  \qquad  & 0.985&0.999 \qquad & 0.982&0.999  \\ 
    $H\to gg$           \qquad  & 0.810&0.985   \qquad &  0.830&0.986  \qquad  & 0.816&0.982 \qquad & 0.736&0.954  \\ 
    $H\to \gamma\gamma$ \qquad  & 0.997&1.000   \qquad &  0.999&1.000  \qquad  & 1.000&1.000 \qquad & 0.997&1.000  \\
    $H\to ZZ$           \qquad  & 0.650&0.958   \qquad &  0.667&0.960  \qquad  & 0.585&0.947 \qquad & 0.535&0.926  \\
    $H\to W^+W^-$       \qquad  & 0.806&0.981   \qquad &  0.801&0.981  \qquad  & 0.771&0.974 \qquad & 0.632&0.952  \\
    $H\to \gamma Z$     \qquad  & 0.921&0.996   \qquad &  0.936&0.996  \qquad  & 0.910&0.993 \qquad & 0.896&0.993  \\
    \bottomrule 
    \end{tabular} 
    \caption{\label{tab:training} Accuracy (left) and AUCs (right) of four classifiers.  } 
\end{table}

\begin{multicols}{2}
\subsection{An attempt of 39-classification}
The above study shows that the multi-classification is very promising in data analysis, so here a more ambitious case of 39-classification will be tried.  For the signal processes,  considering that $Z$ decay to 4 categories (neglect neutrino decay and $W$ fusion processes up to now), i.e,  $e^+e^-$,  $\mu^+\mu^-$, $\tau^+\tau^-$, and $q\bar{q}$,  and that the Higgs to the same 9 decay modes as the above, so there are 36 signals. For a realistic analysis, the backgrounds must be taken into account, especially the irreducible ones. In the analysis of $e^+e^- \to ZH $ study, the irreducible backgrounds mainly come from the SM process of $e^+e^- \to ZZ$. The background can be categorized into 3 classes depending on the decays of $Z$ bosons, i.e,  pure leptonic ($ZZ_{l}$), semi-leptonic($ZZ_{sl}$), and hadronic ($ZZ_{h}$) decays. In conclusion it is a 39-classification problem.  

Same data sets of the signal and extra 3 background processes are pre-processed with the same procedure, which has 39$\times$400 000= 15 600 000 events, which is very challenging  because of memory usage. So we switch to anther deep learning framework, ParticleNet/Weaver\cite{particlenet,weaver}, which has more flexible memory strategy. 

\begin{figure}[htbp]
\centering 
\includegraphics[width=0.8\textwidth,origin=c,angle=0]{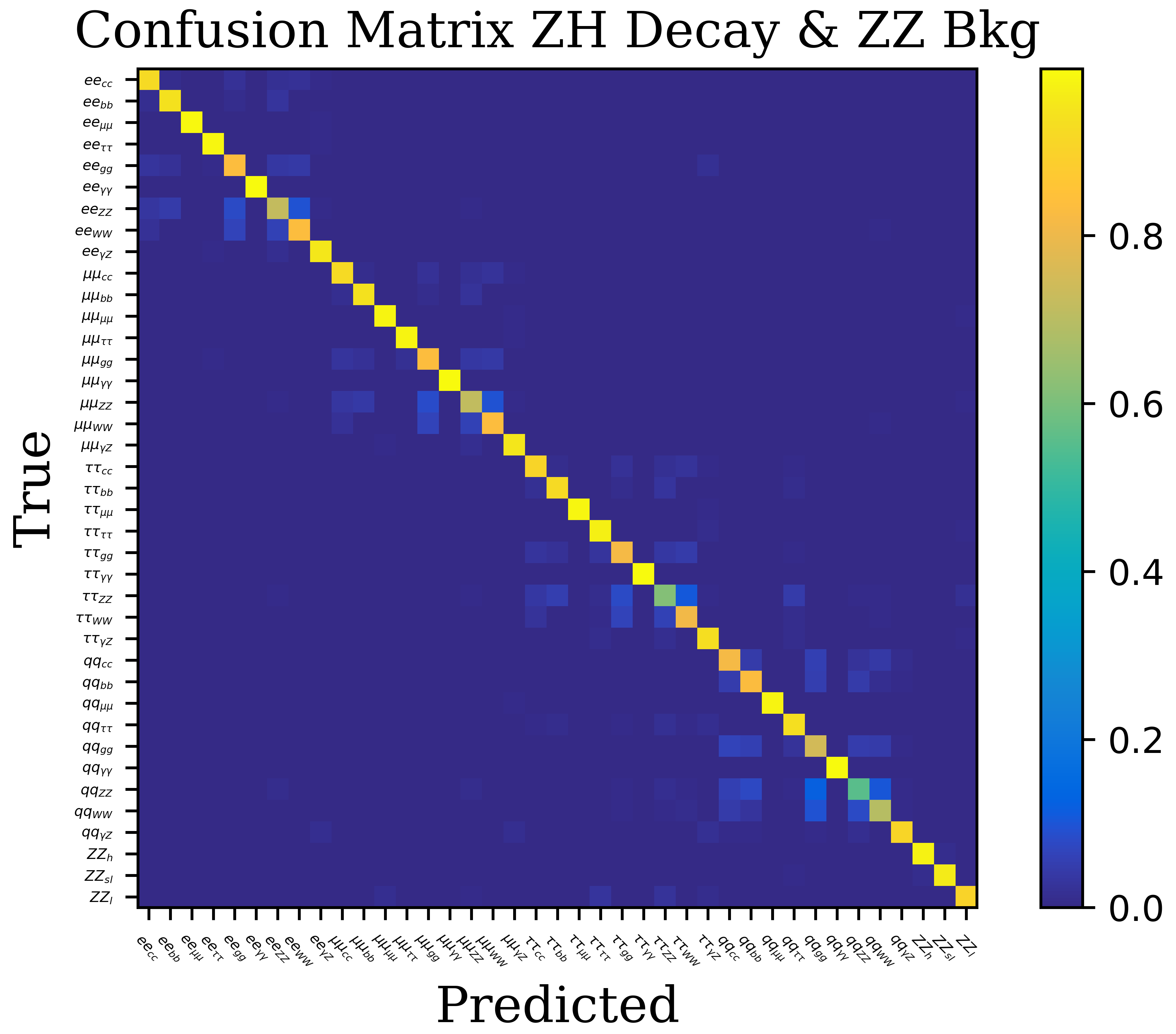}
\caption{\label{fig:conf-39} Confusion matrix of the 39-classification }
\end{figure}

The confusion matrix of the 39-classification is presented in Fig.~\ref{fig:conf-39}, which shows very good separation power among all 39 processes. For the signal, four blocks of the $e^+e^-H$, $\mu^+\mu^-H$, $\tau^+\tau^-H$, and $q\bar{q}H$ processes can be seen clearly, which demonstrate similar patterns as the corresponding ones in Fig.~\ref{confusion}. In each sub-matrix, the $H\to \gamma\gamma$, $\mu^+\mu^-$, and $\tau^+\tau^-$ decays achieve the best performances. Among the four ``blocks'', the mis-classification rates are rather small.  And the $H\to ZZ$ decay doesn't achieve sufficiently good performance as the other decays, which is also consistent with those of the 9-classification.  For the irreducible backgrounds of $e^+e^\to ZZ$, all of three processes are labeled correctly with very high accuracy, greater than 90\%, which indicates that the kinematics of different events can be learnt to discriminate the irreducible backgrounds by the ParticleNet. 

{

In this 39-classification, all 9 Higgs decays in 4 tagging modes with the irreducible backgrounds together can be classified with rather good accuracy. It is different with only one single tagging mode, which indicates the the Higgs decays can be determined with a combined way and much more information. So better precision could be expected naturally.

}

\end{multicols}

\begin{multicols}{2}

\section{Summary and discussion}

In this paper, we presented a study of the classification of the Higgs decays with the state-of-the-art ML approaches at electron-positron colliders. It deploys the ML techniques and try to classify both the signal and background events with only particle level information and obtain the confusion matrices, which can be used in the further data analysis.  This approach is the basis of an efficient and balanced ``one-shop'' analysis, which is possible to measure all Higgs couplings using all detector information and taking all the commonalities and correlations into account. For the analyses of tens or hundreds of channels, they can be repeated using this technique in a few days if all data samples are ready. In contrast, the time could be considerably longer using the conventional analysis method.    

To be frank, this work is only a feasibility study. There are various possibilities to improve and further validate it. One is to enhance the performance by taking the sequential decays of $W$ and $Z$ bosons into account, which can adopt more information and enhance the classification performance. Another endeavor with more physical significance is incorporating some physics processes beyond the SM in the analysis, such as the invisible and semi-invisible decays of the Higgs boson, which can enhance the sensitivity of an experiment to new physics. It is also very constructive to take the full SM backgrounds and main systematic uncertainties into account.

\section*{Acknowledge}
The authors present special thanks to Yunxuan Song, Congqiao Li, Dr. Yu Bai, and Dr. Huilin Qu for useful discussion and advice. The authors thank the IHEP Computing Center for its firm support. This work is supported by the National Natural Science Foundation of China (NSFC) under Contracts Nos. 12075271 and 12047569.

\end{multicols}

\end{CJK*}

\end{document}